\documentstyle[floats,aps,psfig]{revtex}
\begin{document}

\flushbottom

\twocolumn[\hsize\textwidth\columnwidth\hsize\csname @twocolumnfalse\endcsname

\title{Strong Charge Fluctuations in the Single-Electron Box: \\
A Quantum Monte Carlo Analysis}
\author{Carlos P. Herrero,$^{1,2}$ Gerd Sch\"on,$^{1}$ and Andrei D.
Zaikin$^{1,3}$ }
\address{$^1$Institut f\"ur Theoretische Festk\"orperphysik,
	 Universit{\"a}t Karlsruhe, D-76128 Karlsruhe, Germany \\
	 $^2$Instituto de Ciencia de Materiales, C.S.I.C., 
	 Cantoblanco, 28049 Madrid, Spain \\
	 $^3$Lebedev Physics Institute, Leninski pr.\,53,
	 117924 Moscow, Russia}
\date{\today}
\maketitle

\begin{abstract}
We study strong electron tunneling in the
 single-electron box, a small metallic island coupled to an
electrode by a tunnel junction, by means of quantum
Monte Carlo simulations.
We obtain results, at arbitrary tunneling strength,
for the free energy of this system and 
the average charge on the island as a function
of an external bias voltage.
In much of the parameter range an extrapolation to the ground state
is possible.
Our results for the effective charging energy for strong
tunneling are compared to earlier -- in part controversial --
theoretical predictions and Monte Carlo simulations.\\

\end{abstract}
\vskip2pc]

\narrowtext

\section{Introduction}

In the last few years charging effects in low capacitance tunnel junctions
have been studied extensively, on one hand because of their possible
practical applications in single-electron devices, on the other hand because 
of interesting theoretical questions related to them.
These mesoscopic tunnel 
junctions are generic examples of macroscopic quantum systems
with discrete charge states and dissipation \cite{sc90,gr92,so97}. 
They can be fabricated by modern lithographic techniques
with junction capacitances  as low as $\sim 10^{-15}$ -- $10^{-16}$ F.
In these systems single-electron tunneling (SET) is strongly influenced
by the Coulomb interaction  at temperatures $T < E_C/k_B \sim 1$ -- $10 $ K.
The simplest device displaying these effects is the so-called
single-electron box, where a metallic island is
connected via a capacitor $C_G$ and a tunnel junction with
capacitance $C_J$ to a gate voltage source $V_G$ \cite{la91}.
This external voltage polarizes the system,
which allows a continuous tuning of the  properties of the box.
The (bare) charging energy of the system is given by a set of parabolas
$E(n_G, n) = E_C (n_G -n)^2$ as a function of the {\sl integer} number, 
 $n$, of excess electrons in the island and the {\sl continuous} 
dimensionless gate charge $n_G = C_G V_G / e$.
The energy scale and curvature at
small $n_G$ is given by $E_C = e^2/2 C$. It depends on the 
 capacitance of the island $C = C_G + C_J$, and $e$, the electronic charge.
The charging energy is minimized if $n$ is as close to $n_G$ as
possible. Hence at low temperatures,  $k_B T \ll E_C$,
 and very weak tunneling in 
high resistance junctions, $R_t \gg R_K \equiv h /e^2$,
$n$ increases in a stepwise fashion as $n_G$ is increased.
In the range $-1/2 \leq n_G \leq 1/2$, which  for definiteness 
we consider in the
following, the ground state corresponds to $n=0$,
i.e.\ $E_0(n_G)=E(n_G,0)$. 
At finite temperatures higher charge states are excited,
and the expectation value
$\langle n\rangle_T$ has smeared steps, approaching a linear
$n_G$-dependence at high $T$.

In the weak tunneling limit one can proceed perturbatively 
in an expansion in the dimensionless junction conductance  
\begin{equation}
\alpha_t = \frac{1}{4 \pi^2}\, \frac{R_K}{R_t} \; .
\label{alphat}
\end{equation}
In lowest order, transport by sequential single-electron 
tunneling processes is described by the 'orthodox theory' \cite{Ave-Lik}.
On the other hand, the perturbation theory also shows that 
equilibrium properties, such as the ground-state
energy $E_0(n_G)$ and the low-temperature expectation value of the charge
 $\langle n \rangle = n_G - 1/(2E_C) \; dE_0(n_G)/dn_G$ 
get renormalized by electron tunneling processes. 
The latter is observable as a weakening of the
Coulomb blockade  \cite{saclay,ch98}. 
In  first order in $\alpha_t$ one finds for the two quantities
\cite{sc90,gu86,gl90,matveev,go94}
\begin{equation}
E_C^* \equiv \frac{1}{2}  
\left.
\frac{d^2}{dn_G^2}E_0(n_G) \right|_{n_G =0}
 = E_C \left( 1 - 4\alpha_t + O(\alpha_t^2)\right)
\label{perttriv2}
\end{equation}
and (for $-1/2 < n_G < 1/2$)
\begin{equation}
\langle n\rangle = \alpha_t\ln \left(\frac{1+2n_G}{1-2n_G}\right) \; .
\label{perttriv}
\end{equation}
We observe that even for weak $\alpha_t$ the corrections are most
pronounced near the degeneracy points $n_G = \pm 1/2$. In fact, at the
degeneracy points and low temperatures
any order perturbation theory fails. 

In order to cover stronger tunneling and/or a closer vicinity of the
degeneracy points various theoretical approaches have been pursued.
Systematic higher order expansions in $\alpha_t$ have been performed
for equilibrium \cite{gr94,go98}, as well as 
transport properties \cite{ko97}. These expansions
describe the system well for weak to intermediate tunneling strength
as long as the degeneracy points are avoided or the temperature is not
too low. Renormalization-group (RG)
techniques \cite{matveev,fa95} as well as 
a partial summation of arbitrary order processes, accounting for 
resonant tunneling
phenomena  \cite{schoell94}, cover the regime of stronger tunneling.
However, these approaches have concentrated
on two adjacent charge states and require specifying 
a high frequency cut-off of the order of the charging energy, thus
introducing an uncertainty which prohibits a quantitative comparison.
This limitation has been overcome
in the recent work of K\"onig and Schoeller \cite{sk98} within a
`real-time RG' approach. By including higher charge states they obtain
cutoff independent results.
Concentrating on the renormalization of the density matrix they can
further cover nonequilibrium and transport properties. 
A cutoff-independent expression 
for $E_0(n_G)$ had also been obtained
within a `non-crossing' approximation scheme \cite{go94}.

When $R_t$ is lower, closer to $R_K$, the electron
tunneling leads to strong fluctuations of the charge on the island
for all values of $n_G$.
In the limit of high tunneling conductances, 
$1/\alpha_t$ can be treated as a small parameter and nonperturbative 
effects emerge. Also in this regime a RG analysis has been formulated 
\cite{gu86}. Instanton techniques have been developed for 
the equilibrium properties  \cite{ZP88,pa91}, as well as a
real-time saddle point expansion for transport properties \cite{go92}. 
All these approaches \cite{pa91,go92,za93,wa96a,ho97} arrive at the conclusion
that in the strong tunneling regime the  charging energy
$E_C^*$ is renormalized as 
\begin{equation}
E_C^* = f(\alpha_t)E_C \exp(- 2 \pi^2 \alpha_t) 
 \; .
\label{EC*}
\end{equation}
While  there exist general consensus about the exponential
dependence of $E_C^*$ on $\alpha_t$, much controversy remains about 
the pre-exponential factor. The instanton
analysis of Ref. \cite{pa91} yields $f(\alpha_t) \propto \alpha_t$ for
$E_C^* <k_B T \ll E_C$ (which is confirmed 
by the quasiclassical approach \cite{go92})
and $f(\alpha_t) \propto \alpha_t^2$ in the limit $k_B T \ll E_C^*$.
In contrast, in Ref. \cite{wa96a} a cubic dependence  $f(\alpha_t) \propto
\alpha_t^3$ has been found, while Ref. \cite{ho97} reports a linear
dependence for all temperatures. This controversy is one of the
 motivations for us to revisit the problem.

The expression (\ref{EC*}) demonstrates that values of $\alpha_t \gtrsim 0.1$ 
already correspond to  strong tunneling, resulting in a
substantial renormalization. (Accordingly, a dimensionless conductance
with different numerical coefficients, more
 suitable for the strong tunneling regime
could be defined.  To avoid confusion, we prefer to
use the weak-tunneling expansion parameter, as defined in Eq.\,(\ref{alphat}),
throughout this paper.)
In principle Coulomb blockade effects survive at $T=0$ 
for any value of $\alpha_t$.
For instance, clear signs of Coulomb
blockade have been observed for $\alpha_t \sim 0.84$ 
\cite{ch98}. But the experimental observation
of such effects for even larger $\alpha_t$
requires exponentially low temperatures $k_B T \lesssim E_C^*$, and therefore
is hard to achieve. 
Thus, it is of most practical interest to investigate electron
tunneling for values of $\alpha_t$ of order 1 and less. 
For this purpose, and in order to fill the gap between the limits
covered by the analytic approaches, we
analyze the problem numerically by Monte Carlo (MC) techniques.

MC results for the renormalized charging
energy $E_C^*$ had been obtained before by different groups 
 \cite{fa95,ho97,wa97}, and
in the limit of small $\alpha_t$ there is a very good 
agreement among the recent data \cite{ho97,wa97}
and those of perturbation theory \cite{gr94}. However,
deviations arise for $\alpha_t \gtrsim 0.2$. They 
increase with increasing $\alpha_t$ and have a {\it systematic} character, 
hard to explain within the error bars. 
One possible source for the discrepancy for large $\alpha_t \gtrsim
0.4$ could be the fact that  the simulations of Ref. 
\onlinecite{wa97} were done down to substantially lower temperatures 
(typically $k_B T \gtrsim 2\times 10^{-3}E_C$) than those of Ref. \cite{ho97}
($k_B T \gtrsim 10^{-2}E_C$). But this does not account for
the differences at smaller $\alpha_t \lesssim 0.4$ where  the 
temperature $k_B T \sim 10^{-2}E_C$ was found to be sufficient for
convergence. 
The published MC data are also not sufficient to resolve
the discrepancy concerning the prefactor in Eq.\,(\ref{EC*}). 


Other physical  quantities of interest are the ground state energy  of
the system $E_0(n_G)$ (i.e.\ the lowest energy band) 
and the average charge on the island, $\langle n \rangle$,
as a function of the gate charge $n_G$
for general values of the tunneling strength $\alpha_t$.
These quantities have been studied   
analytically in the limits of weak \cite{matveev,go94,schoell94}
and very strong tunneling \cite{fa95,pa91},  at finite (not too low) 
temperatures \cite{gz96,go97}, as well as by the real-time RG analysis of
Ref.~\onlinecite{sk98} which appears to cover a larger parameter range.    
It is clearly  of interest to extend these investigations 
to intermediate values of $\alpha_t$
and low temperature, and to control the analytic results 
 by  numerical means.

This paper is devoted to a detailed MC analysis of the 
single-electron box in the
regimes of weak to strong tunneling, $0 \le \alpha_t
\lesssim 1$. In Sec.\,II, we describe the
model and the computational method employed in our calculations.
The MC results for the renormalized charging energy
at low and finite temperatures are presented in Sec.\,III.
In Sec.\,IV we present 
the free energy $F(n_G)$ as a function of the
gate voltage, and we show results for
the mean charge on the island. A discussion of our results and 
of their relation to other work is given in Sec.\,V.

\section{Model and computational method}

\subsection{Basic formalism}

The grand partition function of the single-electron box
can be 
represented in terms of the path integral\cite{sc90}
\begin{eqnarray}
  Z(n_G,\alpha_t) = \int d\varphi_0 \sum_{m=-\infty}^{\infty}
  \exp(2 \pi i m n_G) \, \times    \nonumber  \\
   \int_{\varphi_0}^{\varphi_0+2\pi m} {\cal D} \varphi \,
   \exp \left( - S[\varphi] \right)
   \hspace{.2cm} .
   \label{part1}
\end{eqnarray}
Here the ``phase'' variable $\varphi (\tau )$ is conjugate to
the island charge, and $m$ is the ``winding'' number of the compact
variable $\varphi$.
The effective action $S[\varphi]$ is given by 
\begin{eqnarray}
     S[\varphi] & = &  \frac{1}{4 E_C}
	   \int_0^{\beta} \left( \frac{d \varphi}{d\tau} \right)^2  d\tau
           \,  +   \nonumber     \\
       & + &  \,   2 \int_0^{\beta} d\tau \int_0^{\beta} d\tau'
      \alpha(\tau - \tau')  \sin^2 \left[ \frac {\varphi(\tau) - \varphi(\tau')}
       {2}  \right],
  \label{action}
\end{eqnarray}
where the first and the second terms account for the charging
energy and the electron tunneling, respectively. 
The kernel $\alpha(\tau)$ reads in Fourier representation 
\begin{equation}
    \alpha(\tau) = - \frac{\pi}{\beta} \, \alpha_t \, \sum_n |\omega_n|
		      \exp(i \tau \omega_n)
       \hspace{.2cm} ,
   \label{kernel}
\end{equation}
for Matsubara frequencies smaller than the electronic bandwidth,
 $|\omega_n| \ll D$.
For $D \gg k_B T$, which will be considered here, the kernel takes the form
$  \alpha(\tau) = \alpha_t \,  ( \pi k_B T )^2 /
	  \sin^2 ( \pi k_B T \tau ) $.
Apart from the bare charging energy scale $E_C$,
the dimensionless conductance $\alpha_t$, and the temperature, 
the model depends on the gate charge $n_G$, which is proportional to
 the applied gate voltage.

The effective action and partition function can be rewritten
 in terms of the phase fluctuations 
$  \theta(\tau) = \varphi(\tau) - 2 \pi m \tau / \beta $, with boundary condition
$\theta(0) = \theta(\beta),$ in the form
\begin{equation}
Z = \sum_{m=-\infty}^{\infty}  
	    \exp(2 \pi i m n_G) \,  I_m(\alpha_t,\beta) \hspace{.2cm} .
   \label{part2}
\end{equation}
The coefficients $I_m(\alpha_t,\beta) =\int {\cal D} \theta  \, \exp \left( -
S_m[\theta ] \right) $ are to be evaluated with the effective action
$ S_m[\theta(\tau)] = S[\theta(\tau) + 2 \pi m \tau / \beta] $. 
They depend on the winding number
$m$, the temperature, and the tunneling conductance, 
but are independent of the gate charge $n_G$.
Thus, from a computational point of view,
the problem reduces to calculating the relative values of
$I_m(\alpha_t,\beta)$, which can be obtained from the 
Monte Carlo simulations apart from an overall normalization constant.
Since the partition function is even and periodic with respect to $n_G$,
$Z(n_G) = Z(-n_G) = Z(n_G + 1)$, 
we can restrict our discussion to the region $0 \le n_G \le 0.5$.

For low tunneling conductance, the island charge is quantized
and at zero temperature, the average number of excess electrons in the box,
$\langle n \rangle$,
is a staircase function of the external voltage.
In general,  at finite temperatures and arbitrary tunneling 
the average charge can be determined from the free energy  $F = - k_B
T \ln Z$ of the island by
\begin{equation}
 \langle n \rangle = n_G - \frac{1}{2 E_C} \, \frac{\partial F}{\partial n_G}
  \hspace{.2cm} .
   \label{mcharge}
\end{equation}
 From the gate-voltage dependence of the free energy one can, further,
define a temperature-dependent 
effective charging energy for the single-electron box as 
\begin{equation}
   E_C^*(T) =  \frac{1}{2}  \left. 
\frac{\partial^2 F}{\partial n_G^2} \right|_{n_G=0}
  =  E_C \left( 1 - \left. \frac{\partial \langle n \rangle}
	   {\partial n_G} \right|_{n_G=0} \right)         \hspace{.2cm} .
   \label{ec1}
\end{equation}	
The zero-temperature limit of this  quantity, $E_C^* \equiv E_C^*(0)$, 
is the renormalized charging energy discussed in the introduction. It 
coincides with the classical 
energy scale $E_C$ for weak tunneling ($\alpha_t \ll 1$), but is
renormalized in systems with stronger tunneling conductance.
At high temperatures,  the free energy $F(n_G)$ depends weakly on
$n_G$, and the curvature $E_C^*(T)$ approaches zero.
By using Eqs.\,(\ref{part2}) and (\ref{ec1}), 
this effective charging energy $E_C^*(T)$ can also be expressed as 
\begin{equation}
E_C^*(T) =
2 \pi^2 k_B T \langle m^2 \rangle_{n_G=0} \;,
   \label{ec2}
\end{equation}
where $\langle m^2 \rangle_{n_G=0}$ is a moment of the coefficients
$I_m(\alpha_t,\beta)$.

\subsection{Monte Carlo method}

MC simulations have been carried out by the standard
discretization of the quantum
paths into $N$ (Trotter number) imaginary-time slices \cite{su93}.
In order to keep roughly the same precision in the calculated
 quantities,  as the temperature is reduced, the number of time-slices
$N$ has to increase as $1 / T$.
 We have found that a value $N = 4 \beta E_C$ is sufficient to
 reach convergence of $I_m$, even for the lowest studied
 temperatures and for strong tunneling, where convergence has
 been reported to be slower \cite{wa97}.
 Thus, the imaginary-time step employed in the discretization of the
 paths is $\Delta \tau = \beta / N \sim 1 / (4 E_C) $.  This means
 that the high-energy cut-off $\omega_c$ associated with this 
discretization is $\omega_c \sim 2\pi /\Delta \tau \approx 25E_C$.
Repeating the calculation for a few data
points with  higher values of $N$ and  high-energy cut-off
did not change our results.

The classical Metropolis MC sampling \cite{bi88}
has been used to obtain finite-temperature results, 
and to extrapolate, where possible, 
to $T = 0$. The partition function has been sampled
according to Eq.\,(\ref{part2}) for temperatures down to
$k_B T = E_C / 500$. These low temperatures were
necessary to determine the zero-temperature
effective charging energy for large tunneling conductance
(see below).
A simulation run proceeds via successive MC steps (MCS).
In each MCS, all path-coordinates are updated.
At each studied temperature, the maximum distance allowed
for random moves was fixed in order to obtain an acceptance ratio
of about $50 \%$.  For each set of parameters ($\alpha_t$, $T$),
we generated $\sim 3\times10^5$ quantum paths for the calculation of
ensemble-averaged values. The starting configuration for
the MC runs was taken after system equilibration at the
considered temperature. In general,
equilibration runs of about  $2 \times10^4$ MCS were sufficient,
but in some extreme cases, especially for strong tunneling
conductance, equilibration runs of about $1 \times10^5$ MCS were
necessary.

In some cases, the acceptance ratio for jumps between different
winding numbers during a MC run becomes very low, and sampling
by direct jumps is inefficient. This happens, in particular, at high 
temperatures ($k_B T > E_C$), where $I_m/I_0$ is very small
for $m \ne 0$, and at low temperatures (especially for strong
tunneling), where the most relevant phase paths for different
winding numbers are very different.
In these cases, we have calculated $I_{m'}$ by carrying
out simulations for fixed winding number (say $m$), and evaluating
in this fixed-$m$ ensemble the average value 
\begin{equation}
   R_{mm'} = \langle  \exp \left( S_{m}[\theta] -
	      S_{m'}[\theta] \right) \rangle_{m}
\end{equation}
for $m \ne m'$.
The average value $R_{mm'}$ so defined coincides with the ratio
$I_{m'}/ I_m$, since by definition we have
\begin{equation}
    \frac{I_{m'}}{I_{m}} = \frac {\int {\cal D} \theta \, \exp 
            \left( - S_{m}[\theta] \right)
	  \exp \left( S_{m}[\theta] - S_{m'}[\theta] \right) }
	  { \int {\cal D} \theta \, \exp \left( - S_{m}[\theta] \right)  }
    \hspace{.2cm} .
\end{equation}
By definition we have $R_{m'm} = 1/R_{mm'}$.
We have checked  that the MC simulations satisfy this consistency relation 
within the statistical noise.
We have also checked for some sets of parameters ($T$, $\alpha_t$)
that this method of calculating relative
values of $I_m$ gives the same results as the direct Metropolis method,
in which the winding number changes during a MC run.

The MC method allows us to
calculate the partition function $Z$
as a function of $T$, $\alpha_t$, and $n_G$. There is,
however, a limitation on  the parameter range
that can be studied by this method. As noted in earlier
publications \cite{fa95,wa97},  due to the term $\exp(2 \pi i m n_G)$
in Eq.\,(\ref{part2}), the ratio $Z(n_G)/Z(0)$ for $n_G \neq
0$ rapidly approaches 0, as the temperature is lowered. Since the MC method
provides the coefficients $I_m$ with a limited accuracy, the partition
function $Z(n_G)$ can be determined in a reliable way only when 
the ratio $Z(n_G)/Z(0)$ is larger than the numerical error in $I_m$.
The temperature range where reliable results can
be obtained for a given value of $n_G \ne 0$ reaches to lower values
as  $\alpha_t$ is increased, 
since the larger is $\alpha_t$, the slower $Z(n_G)/Z(0)$ decreases at
low $T$. We found empirically
that the lowest temperature for which the whole range
$0 \le  n_G \le 0.5$ can be studied by the present method 
with sufficient  accuracy,
scales roughly as $k_B T_{min} \sim  E_C^*(0) / 20$.
On the other hand, this limitation does not apply  for $n_G \approx 0$.
Hence, the effective charging energy $E_C^*(T)$ can be evaluated down 
to much lower temperatures.

\section{Effective charging energy}

\subsection{High-temperature regime}

At high temperatures ($k_B T \gg E_C$), 
the partition function of the single-electron box, including the
effect of electron tunneling, can be approximated 
by an expression similar to the classical result \cite{wa96a} 
\begin{eqnarray}
Z  & \simeq & \sum_{n=-\infty}^{\infty}
	 \exp \left[ - \beta \tilde E_C (n - n_G)^2 \right] =  
	 \nonumber  \\
   & = & \sum_{m=-\infty}^{\infty}  
	    \exp \left[2 \pi i m n_G   - \frac {\pi^2 m^2}
	     {\beta \tilde E_C(T)}  \right] 
       \hspace{.2cm} .
  \label{part3}
\end{eqnarray}
The temperature-dependent parameter $\tilde E_C(T)$ takes into account
quantum fluctuations. It reduces to the bare charging energy $E_C$ 
both in the weak tunneling limit and at high temperatures.
In the second form we made use of a Poisson resummation to establish
the relation to the winding number representation. 
At sufficiently high temperatures, where only winding numbers
$m = 0, \pm 1$ 
have a non-negligible contribution ($I_m/I_0 \ll I_1/I_0$ for $|m| > 1$),
$\tilde E_C(T)$ is well defined. In this case
Wang and Grabert \cite{wa96a} expressed the result of 
a semiclassical calculation, where Gaussian fluctuations 
around the classical paths are allowed, in the
form  (\ref{part3}). They find
\begin{equation}
\tilde E_C(T) = \frac{E_C}{1 + 2 \, \alpha_t \, \beta E_C} \; .
   \label{et1}
\end{equation}
As the temperature is lowered, $\tilde E_C(T)$
decreases as quantum fluctuations become more
prominent. At still lower temperatures
 it is not guaranteed that the partition function
can be parameterized in the form (\ref{part3}).

If the partition function is of the form (\ref{part3}) and
 $\tilde E_C(T)$ can be defined, the latter is related to the effective
charging energy  $E_C^*(T)$, Eq.\,(\ref{ec1}), by
\begin{equation}
E_C^*(T) = \frac{4 \pi^2} {\beta} \exp \left[ - \frac {\pi^2}
       {\beta \tilde E_C(T)}  \right] \hspace{.2cm} .
  \label{ecet1}
\end{equation}
This implies that at high temperatures, $E_C^*(T)$ vanishes
proportional to $T \exp [- \pi^2 / (\beta E_C)]$.

\begin{figure}
\centerline{\psfig{figure=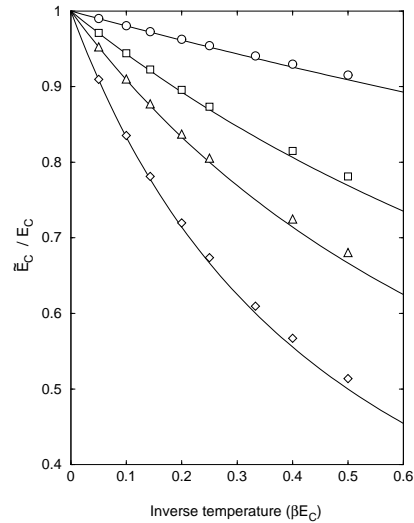,height=7.0cm}}
\caption{
The parameter $\tilde E_C(T)$ is plotted  in the
 high-$T$ region,
 for several values of the tunneling conductance $\alpha_t$. Symbols
 indicate results of the Monte Carlo simulations: $\alpha_t =
 0.1$ (circles), 0.3 (squares), 0.5 (triangles), and 1 (diamonds).
 Error bars are smaller than the symbol size.
 Lines correspond
 to Eq.\,(\protect\ref{et1}), derived from semiclassical calculations in
 Refs.\,~\protect\cite{wa96a,wa96b}.
 }
 \end{figure}

We  determined $\tilde E_C(T)$ by fitting 
the coefficients $I_m$  obtained from the MC simulation  
to a Gaussian profile. At sufficiently high temperatures,
where  $I_m$ converges quickly as a function of $m$, this is obviously
possible. For this range the results are shown in Fig.\,1.
Our  data points (symbols) are compared with the
expression (\ref{et1}) (continuous lines) for several
values of $\alpha_t$. An increase in the  strength of tunneling and
 quantum fluctuations leads to  a decrease of $\tilde E_C$.
We observe good agreement between MC and analytic
results up to $\beta E_C \sim 0.2$, even for the highest junction conductance 
considered.
For larger values of $\beta$ the high-temperature approximation
(\ref{et1}) becomes insufficient, dropping below the MC results.

In the temperature range shown in Fig.\,1 the 
temperature dependence of the effective
charging energy $E_C^*(T)$ is dominated by the exponential
term in Eq.\,(\ref{ecet1}), i.e.\ we find vanishingly low values
for $E_C^*$, irrespective of the strength of the tunneling
conductance.

\subsection{Low-temperature regime}

We now turn to the low-temperature region,  where the 
quantum fluctuations of the island charge dominate the 
thermal fluctuations and are responsible for the
reduction of the effective charging energy $E_C^*(T)$.
The latter
can be conveniently calculated from the mean-square winding number
obtained from MC simulations at finite temperatures, see Eq. (\ref{ec2}). 
Our results are displayed in Fig.\,2 for several
values of the tunneling conductance $\alpha_t = 0, ..., 0.4$. 
In general, $E_C^*(T)$ grows as the
temperature is lowered, and reaches a plateau, the renormalized
charging energy $E_C^*(0)$, at low $T$. 
This saturation of $E_C^*(T)$ should be found for any value of $\alpha_t$.
However, for growing $\alpha_t$ lower and lower
temperatures are required, consistent with the corresponding decrease of
$E_C^*(0)$, which sets the scale for the convergence. 

\begin{figure}
\centerline{\psfig{figure=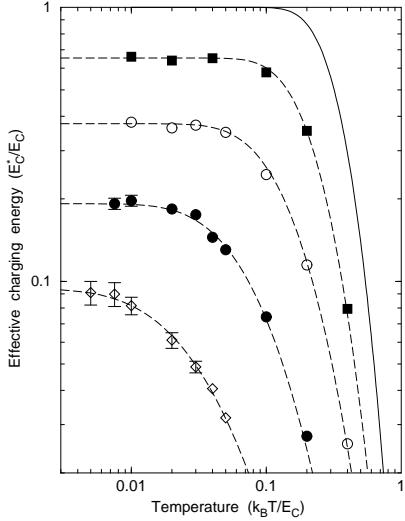,height=7.0cm}}
\caption{
Effective charging energy $E_C^*(T)$ as a function of temperature
for several values of the tunneling conductance. From top
to bottom: $\alpha_t$ = 0 (continuous line), 0.1 (squares),
0.2 (open circles), 0.3 (black circles), and 0.4 (diamonds).
Dashed lines are guides to the eye.
Error bars are of the order of the symbol size, unless shown
explicitly.
}
\end{figure}

At low temperature one can also  employ
an alternative method of calculation of $E_C^*$, and thus provide a
consistency check. To do so, we first
expand the free energy of the system in powers of $n_G$, 
\begin{equation}
  F(n_G,T) = E_C^*(T) \, n_G^2 + A_4(T) \, n_G^4 + A_6(T)\,  n_G^6 + ...
  \label{free}
\end{equation}
Taking a derivative of the partition function in Eq.\,(\ref{part2})
with respect to $n_G$ we find 
\begin{eqnarray}
  A_4(T) & = & \frac {2 \pi^4} {3 \beta} \, \left[ 3 \, \langle m^2 \rangle^2 -
	   \langle m^4 \rangle )  \right]_{n_G=0} 
	   \hspace{.2cm} , 
       \label{coef4}   \\
  A_6(T) & = & \frac {4 \pi^6}{45 \beta} \, \left[ \langle m^6 \rangle -
	   15 \, \langle m^4 \rangle \langle m^2 \rangle +
	   30 \, \langle m^2 \rangle^3 \right]_{n_G=0}  
	   \hspace{.2cm} ,
       \label{coef6}
\end{eqnarray}
and analogous expressions for the higher order terms. Provided that
the low-temperature behavior is regular we can conclude from
Eq.\,(\ref{ec2}) that for $\beta E_C^*(0) \gg 1$, 
the average $\langle m^2 \rangle$
diverges as $\langle m^2 \rangle \propto \beta$.  Similarly, 
one has $\langle m^4 \rangle \propto \beta^2$ and 
$\langle m^6 \rangle \propto \beta^3$ in this limit.  
Since the coefficients $A_n(T)$ are
finite at all temperatures including $T=0$, one obtains 
\begin{equation}
\langle m^4 \rangle / \langle m^2 \rangle^2 = 3 + O(1 / \beta),\;\;\;
\langle m^6 \rangle / \langle m^2 \rangle^3 = 15 + O(1 / \beta),
\label{moments}
\end{equation}
and similar relations for the higher moments of $m$. This implies that
at low $T$ the distribution of the coefficients $I_m$ must be close to 
a Gaussian function of $m$,
up to high winding numbers $|m| \lesssim \beta E_C^*(0)$. 
In other words, in the limit of low $T$ we can rewrite $I_m$ in the form 
\begin{equation}
I_m \propto \exp[ - a_2(T) \, m^2 - a_4(T) \, m^4 - \dots],
\label{Im}
\end{equation}
with $a_2(T) \sim 1 / (\beta E_C)$ and $a_4(T) \sim 1 / (\beta
E_C)^2$. Hence, we have $a_4(T) / a_2(T) \sim 1 / (\beta E_C)$
(the latter relation can be also derived from the fact that
$I_m(T)$ scales at low $T$ as $I_m(T) = I[m/\sqrt{\beta E_C}]$). 
Combining Eqs.\,(\ref{part2}) and  (\ref{Im}) and defining 
$\tilde E_C(T)=\pi^2/\beta a_2(T)$ one obtains in the low-temperature limit
\begin{equation}
 Z  \simeq  \sqrt \frac{\pi}{\beta \tilde E_C(T)}  \sum_{m}
	 \exp \left( 2 \pi i m n_G - \pi^2 m^2 / [\beta \tilde E_C(T)]
	 \right)
	 \hspace{2 mm}  .
   \label{part5}
\end{equation}
This relation is valid only for small values of $n_G$, since
for $n_G \approx 1/2$ the sum converges slowly and higher values of
$m$, for which the $I_m$ are no longer Gaussians,
gain importance.
Note that this expression for the partition function is formally 
identical to the high-temperature form given in Eq.\,(\ref{part3}).

It is easy to show that in the limit $T \to 0$ the two quantities 
$\tilde E_C(0)$ and $E_C^*(0)$ coincide with each other. 
Indeed, in this limit the sum in Eq.\,(\ref{part5})
can be replaced by an integral, giving 
$\langle m^2 \rangle_{n_G=0} = \beta \tilde E_C(0) / (2 \pi^2)$
and the identity $\tilde E_C (0)= E_C^*(0)$ becomes obvious. 
At low but finite $T$ one finds from 
Eq.\,(\ref{part5}) 
\begin{equation}
\frac{E_C^*(T)} {\tilde E_C(T)}  \simeq  1 - 4 \beta \tilde E_C(T)
   \exp [ - \beta \tilde E_C(T) ]  
   \hspace{2mm} .
\label{ecet2}
\end{equation}

The relations described above are well reproduced by our
numerical MC analysis. Consistent with the relation
(\ref{part5}), at low $T$ the coefficients  $I_m$ 
are well described by a single Gaussian function,
up to high values of $m$, characterized by only 
{\it one} parameter $\tilde E_C(T)$.
This parameter is presented in Fig.\,3 as a function of temperature
for $\alpha_t = 0.5$. Also shown are the results for  $E_C^*(T)$.
The data points for  $\tilde E_C(T)$ are indicated by  open squares,
those for the effective charging energy $E_C^*(T)$ by black circles.
Both converge to each other on a temperature scale set by 
$\tilde E_C(T)$, consistent with (\ref{ecet2}).
We also note that both functions converge to a common plateau at low 
temperatures, which defines $E_C^*(0)$.
Similar convergence is found for all other values of $\alpha_t$;
the results for $E_C^*(T)$ and $\tilde E_C(T)$ derived
from our MC simulations are indistinguishable 
(within the statistical noise) for temperatures
lower than $\sim E_C^*(0)/(5 k_B)$.

\begin{figure}
\centerline{\psfig{figure=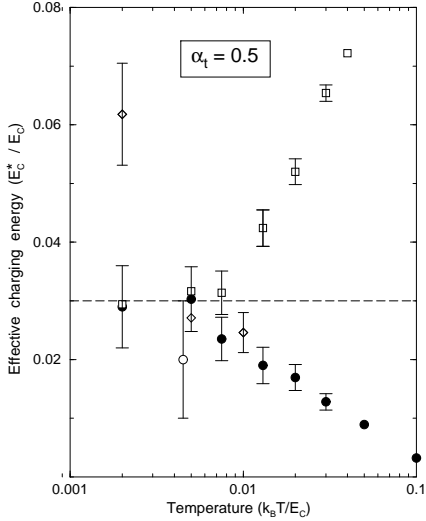,height=7.0cm}}
\caption{
Convergence of the low-temperature effective charging energy
$E_C^*(T)$ as a function of the temperature, for a
tunneling conductance $\alpha_t = 0.5$.
The results of the present MC simulations are represented
by black symbols.
Open squares are results for $\tilde E_C(T)$ derived
from our Monte Carlo simulations.
For comparison, results for $E_C^*(T)$ found in earlier MC
simulations for the
same tunneling conductance are given: Diamonds,
data of Wang {\em et al.}\,~\protect\cite{wa97};
open circle, data point of Hofstetter and Zwerger
~\protect\cite{ho97}. Error bars are shown when they are larger
than the symbol size.
The horizontal dashed line indicates the value to which
our results converge at low temperature.
}
\end{figure}

Clearly, the two methods are not independent from each other.
They are equivalent once the Gaussian distribution
of the coefficient $I_m$ has been established. However, the fact that
at low temperature
the data display the Gaussian distribution with the required accuracy
 provides a consistency check for our MC procedure. The
combination of both approaches  increases
the reliability by reducing the chance
of systematic errors.

\begin{figure}
\centerline{\psfig{figure=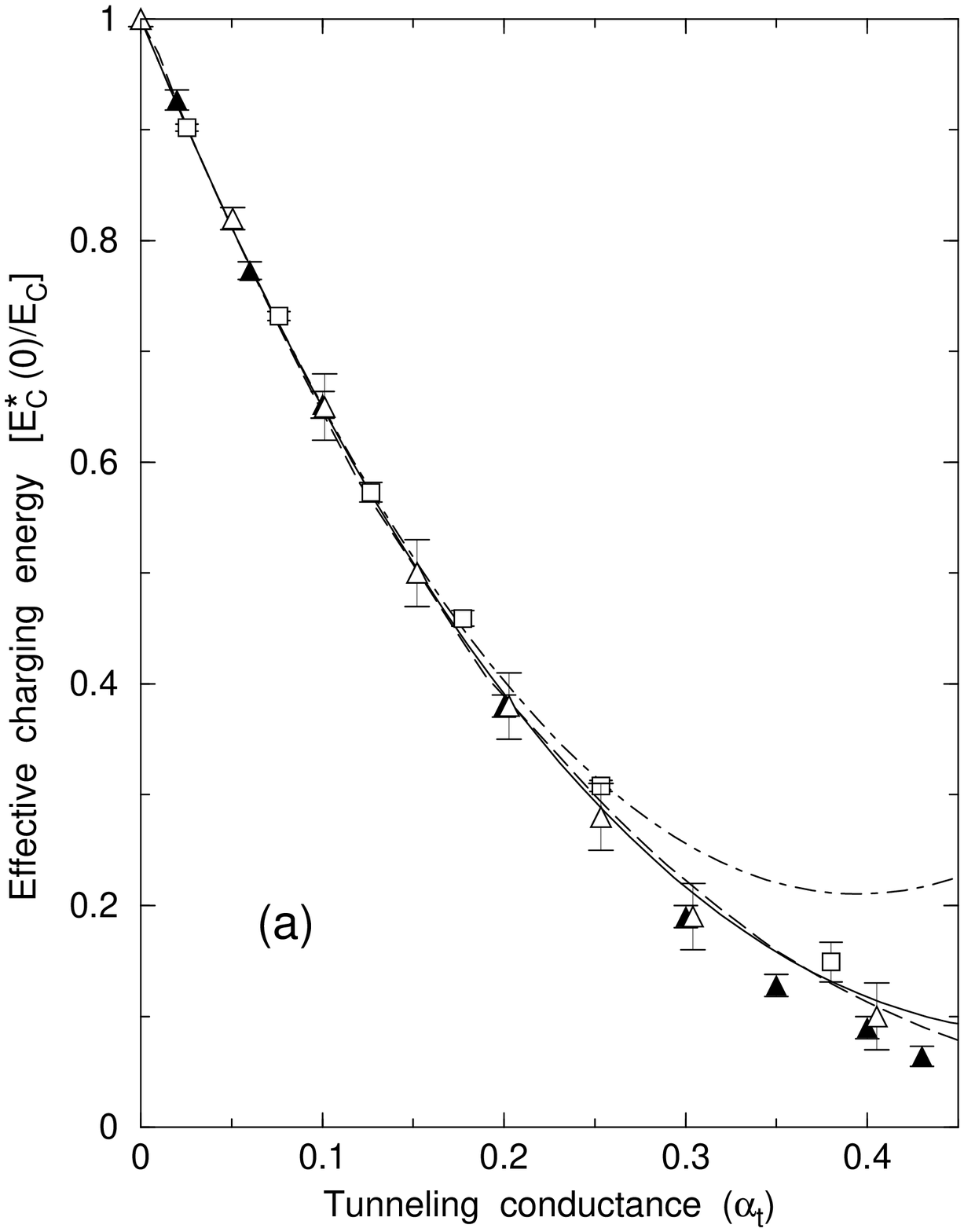,height=8.0cm}}
\vspace*{0.5cm}
\centerline{\psfig{figure=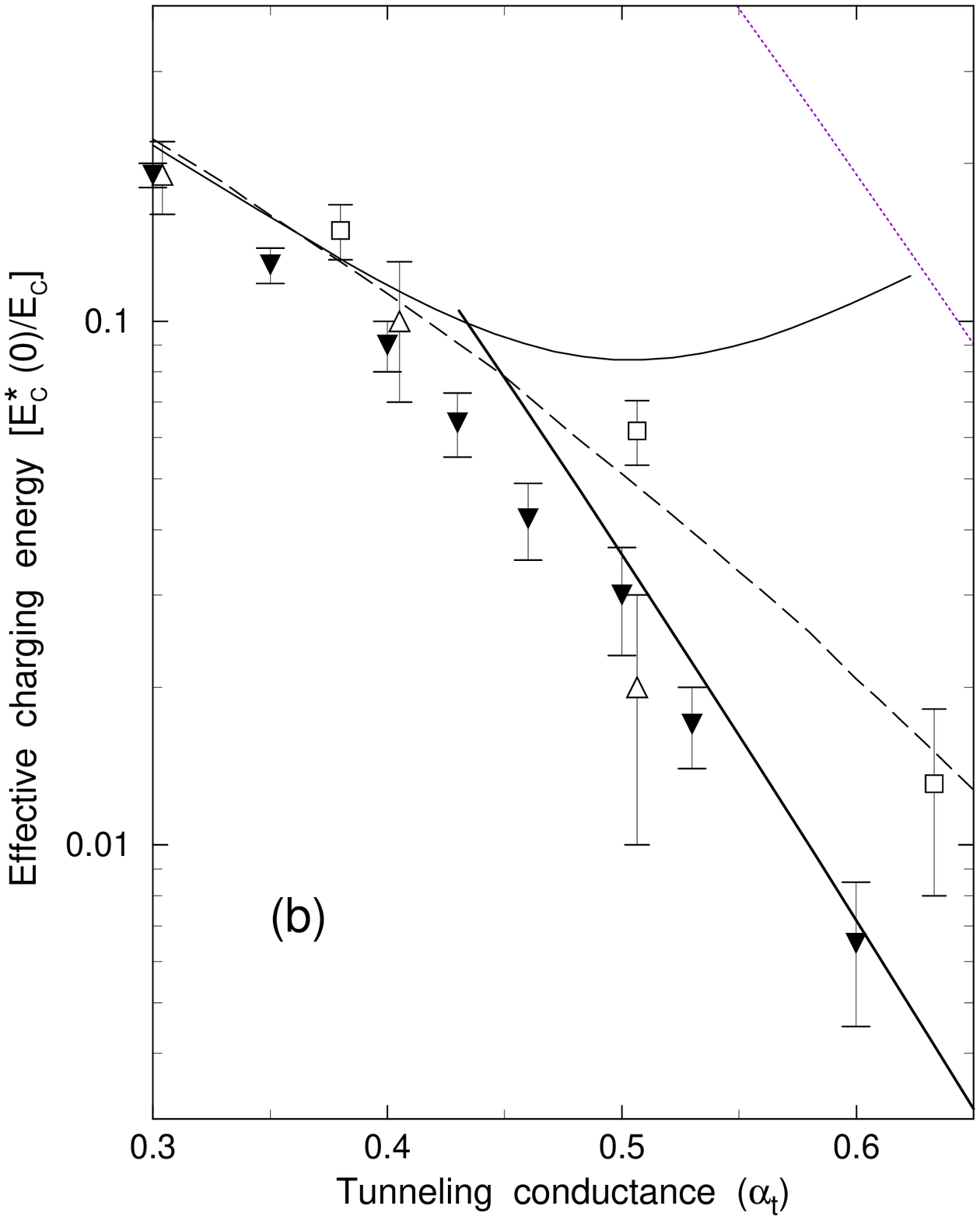,height=8.0cm}}
\vspace*{0.2cm}
\caption{
Low-temperature effective charging energy, $E_C^*(0)$:
(a) in the conductance region
up to $\alpha_t = 0.4$, and (b) for strong tunneling.
Black symbols: results of the present MC simulations.
Data from earlier simulations are represented by
open symbols: triangles, from Ref.\,~\protect\cite{ho97} and
squares from Ref.\,~\protect\cite{wa97}.
The dashed-dotted and continuous lines correspond to second and
third order perturbation theory in $\alpha_t$,
respectively. The dashed line represents the real-time
RG of Ref.~\protect\cite{sk98}.
The bold and dotted lines in (b) correspond
to nonperturbative calculations in Refs.\,~\protect\cite{pa91}
and ~\protect\cite{wa96a}, respectively.
}
\end{figure}

For comparison we also present in Fig.\,3 the 
MC results of two different 
groups for the same tunneling conductance: diamonds \cite{wa97,wa96b}
and open circles \cite{ho97}. All of them
are compatible with those obtained here except for
those found in Ref. \onlinecite{wa97} for $\beta E_C = 500$,
which lie clearly higher than our general trend.

The results obtained for $E_C^*(0)$ from the low-temperature limit
of $E_C^*(T)$ and $\tilde E_C(T)$
are shown in Fig.\,4 as black symbols.
For comparison, we also present analytic and earlier
 MC results.
In Fig.\,4(a), we show the data for low to intermediate
 values of the tunneling 
conductance,  $0 \le \alpha_t \le 0.4$, and 
in part (b) we present the effective charging energy
$E_C^*(0)$ for the strong-tunneling regime on a
semilogarithmic plot.

Our MC results (black symbols) follow closely those 
given by Hofstetter and Zwerger in Ref. \onlinecite{ho97} 
(open triangles), while  for $\alpha_t > 0.3$ the MC data
of Wang {\sl et al.} \cite{wa97} (open squares) are systematically
higher than those found here.
For completeness we add that very recently G\"oppert {\sl et al.}
\cite{go98} carried out MC simulations
 in the charge representation, which implies an expansion
in $\alpha_t$, rather than the phase
representation employed here and in previous MC simulations
\cite{ho97,wa97}. Since no error bars
were given in that work, with symbol sizes too large
to allow a distinction of different results, we do no compare
with our results at this stage.

In the figure we also  present the analytic results of
perturbation expansions in the tunneling conductance 
up to second (dashed-dotted line) and third order
(continuous line).
In third-order, G\"oppert {\em et al.} \cite{go98}
found for the zero-temperature renormalized charging energy
\begin{equation}
\frac{E_C^*(0)}{E_C} = 1 - 4 \, \alpha_t + A_2 \, \alpha_t^2 
	  + A_3 \, \alpha_t^3 + O(\alpha_t^4)
        \hspace{0.2cm}  ,
\label{pert}
\end{equation}
with $A_2$ = 5.066 and $A_3 = -1.457$.
We also show the $\alpha_t$-dependence of the
renormalized charging energy $E_C^*(0)$, as obtained from the
real-time RG calculations  \cite{sk98} (dashed line).
Our numerical values for $E_C^*(0)$ agree with the results of the
perturbation expansion and the RG approach only for $\alpha_t \lesssim 0.25$.

In Fig.\,4(b) we display results  for $E_C^*(0)$ for stronger
tunneling and compare with the predictions of several analytical
calculations. In this regime the weak-tunneling
expansions fail, while the real-time RG results  \cite{sk98}
(dashed line) show a qualitatively correct trend
throughout. Already for $\alpha_t \gtrsim 0.5$ our MC results
get close (within the error bars) to the result obtained 
by a nonperturbative instanton calculation \cite{pa91} (bold line)
\begin{equation}
E_C^*(0) =  16 \pi^4 \alpha_t^2 E_C \exp(-2 \pi^2 \alpha_t + \gamma)
   \hspace{.2cm} ,
\label{inst}
\end{equation}
where $\gamma$ is Euler's constant.
Since the numerics confirms the  result (\ref{inst}) 
already  for $\alpha_t \sim  0.5$, and since the accuracy of the
instanton analysis \cite{pa91} should increase with  $\alpha_t$,
we expect that (\ref{inst}) remains accurate also in the regime
$\alpha_t >0.6$ not covered by our MC simulations.
For comparison, we also present in Fig.\,4(b)
the result for $E_C^*(0)$ found in Ref. \onlinecite{wa96a}
for strong tunneling at $T = 0$ (dotted line),
which predicts a pre-exponential factor  $\sim \alpha_t^3$. It
lies more than one order of magnitude
higher than the results of our MC simulations.

\section{Gate-charge dependence}
\subsection{Free-energy band}

The MC method employed here allows us to determine 
the ''free-energy band" $F(n_G)$ as a function of the gate voltage, $n_G$,
at finite temperatures.
As an example, we display in Fig.\,5(a)
results  at different temperatures for a dimensionless
conductance $\alpha_t$ = 0.3.
The error bars of the Monte Carlo simulations associated 
with the free energy are largest around $n_G = 0.5$.
At high temperatures,  $F(n_G)$ approaches a cosine shape, 
$F(n_G) \simeq E_C^*(T) [1 - \cos(2 \pi n_G)]/(2 \pi^2)$,
as expected for the classical limit. 
At low temperatures, the free-energy band is closer to a parabolic
shape, which is what we expect in
limit of vanishing junction conductance $\alpha_t \ll 1$, with
deviations which are most pronounced near $n_G = 1/2$.

\begin{figure}
\centerline{\psfig{figure=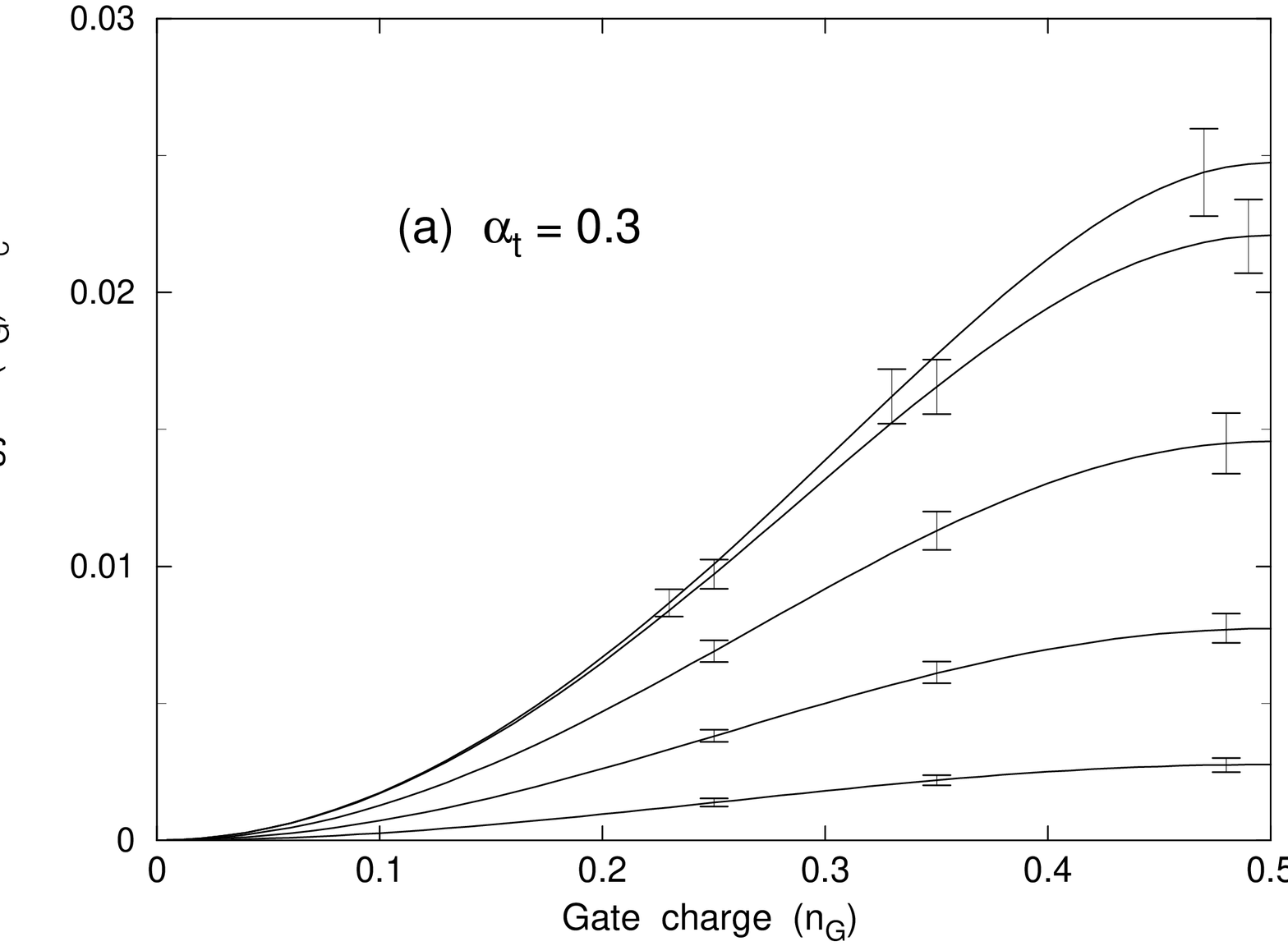,height=7.0cm}}
\vspace*{-1.5cm}
\centerline{\psfig{figure=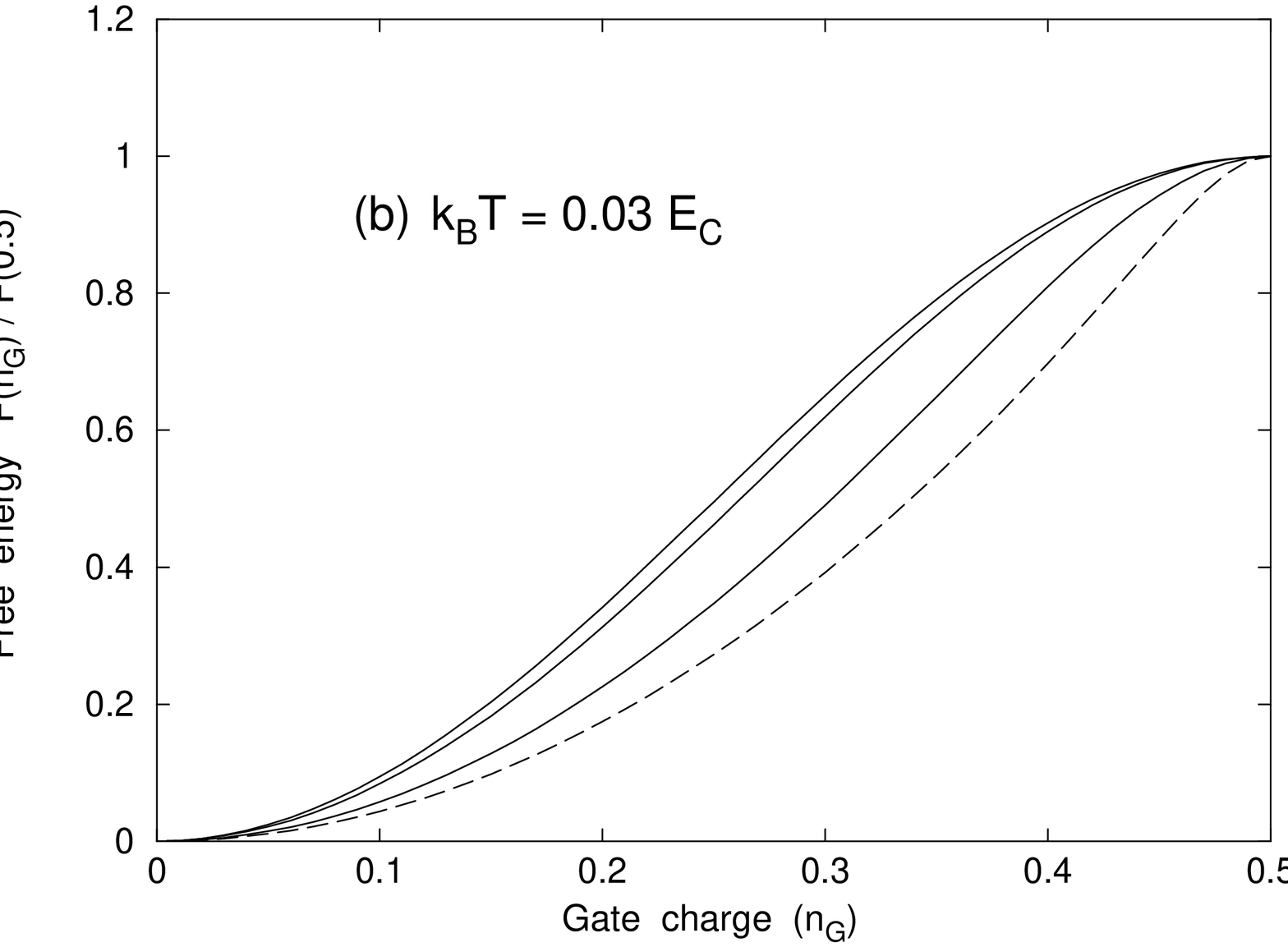,height=7.0cm}}
\vspace*{-1cm}
\caption{
Free energy vs. the dimensionless gate voltage, $n_G$
as obtained from Monte Carlo simulations.  (a)
Data for  junction conductance $\alpha_t = 0.3$
and different temperatures. From top to
bottom, $k_B T / E_C =$ 0.01, 0.03, 0.05, 0.1, and 0.2.
The width of the free-energy
band, $F(n_G$=$0.5) - F(n_G$=$0)$,
decreases as the temperature increases.
(b) Same at fixed temperature $k_B T = 0.03 E_C$,
for several values of the tunneling conductance. From top to
bottom: $\alpha_t$ = 0.5, 0.3, 0.1, and 0 (dashed line).
For each $\alpha_t$, the free energy has been normalized by
its value at $n_G$ = 0.5.
}
\end{figure}

A complementary picture is provided in  Fig.\,5(b). Here $F(n_G)$
is plotted for fixed temperature, $k_B T = 0.03 E_C$, while
$\alpha_t$ is varied. 
The band is close to a cosine function for large conductance,
and clearly differs from this shape for small $\alpha_t$.
For the ease of comparison in this figure the free energy has been
normalized for each $\alpha_t$ to its value at $n_G = 0.5$.

\begin{figure}
\centerline{\psfig{figure=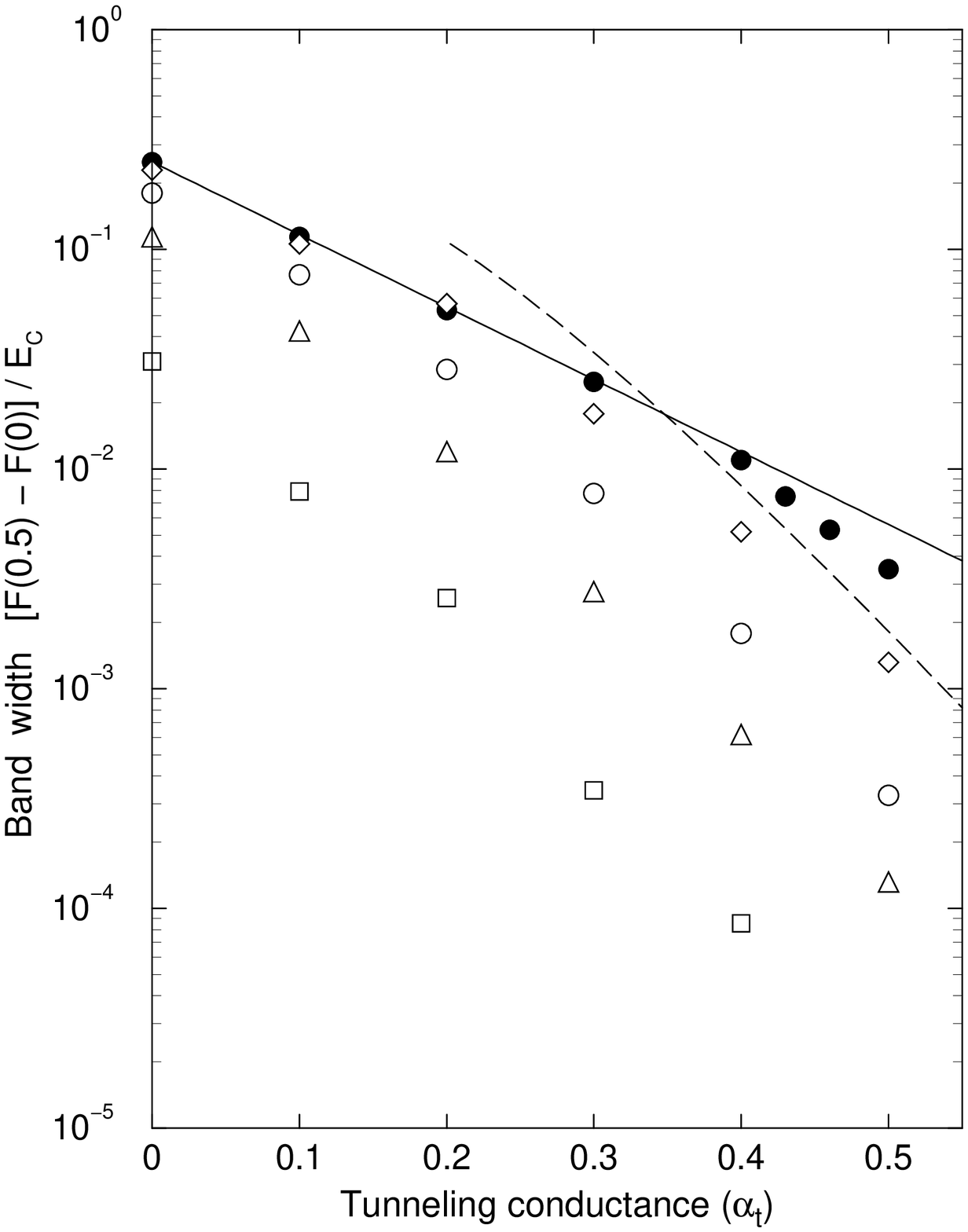,height=7.0cm}}
\caption{
Width of the free-energy band, $W = F(n_G$=$0.5) - F(n_G$=$0)$,
as a function of the junction conductance $\alpha_t$
at several temperatures. Open symbols are finite-temperature results
for $k_B T / E_C =$ 0.4 (squares), 0.2 (triangles),
0.1 (circles), and 0.03 (diamonds). Black symbols correspond
to the lowest temperature reached in our MC simulations.
The continuous line is a fit to the expression
$W = W_0 \exp(- K \alpha_t)$, with $K = 7.6 \pm 0.2$.
The dashed line is the prediction for the low-temperature
band-width for strong-tunneling \protect\cite{pa91}.
}
\end{figure}

The temperature dependence of the  bandwidth
$W(T) = F(n_G$=$0.5) - F(n_G$=$0)$  is shown in
Fig.\,6 for several values of $\alpha_t$.
 For $k_B T > E_C$, one has
$W(T) \simeq E_C^*(T) / \pi^2 \simeq 4 k_B T
\exp[ - \pi^2 / (\beta E_C)]$, as follows from
the high-temperature expression for $E_C^*(T)$ given
in Eq.\,(\ref{ecet1}).
In general, one finds that $W(T)$  increases as the temperature 
decreases, converging to a low-temperature limit. 
But this convergence is slower than
that found for the renormalized charging energy 
$E_C^*(0)$.
The black symbols in Fig.\,6 correspond
to the lowest temperature we could study for each $\alpha_t$
value. Within the 
 temperature range analyzed here we reached
saturation of $W(T)$ only for $\alpha_t \lesssim 0.3$.
In this range  the low-temperature results can be fitted to an
exponential law of the form $W(T=0) = W_0 \exp(- K \alpha_t)$
where $W_0=E_C/4$ and $K = 7.6 \pm 0.2$ (solid line). 
This is still consistent, within the error bars, with
the expression derived for weak tunneling \cite{go94}
$W=W_0(1-8\alpha_t \ln 2)$.
For strong tunneling, one expects for the bandwidth a similar
scaling as a function of $\alpha_t$ as for the
renormalized charging energy, i.e.\ $W(T=0) \sim \exp(- 2 \pi^2 \alpha_t)$.
This form, with prefactors as predicted by
the instanton calculation of Ref.\, \onlinecite{pa91}, is 
shown by the dashed line in Fig.\,6.
The comparison with our Monte Carlo results reveals differences.
While the explicit result of Ref.\, \onlinecite{pa91} is not supported,
the exponential dependence may be reached at larger values of
$\alpha_t$ than covered so far.

Note that even for weak junction conductance 
(where a good convergence of $W(T)$ to its 
low temperature value is demonstrated), 
the second derivative $F''(n_G)$ close to $n_G = 0.5$ 
is still changing at the lowest temperatures
considered here. This is in line with the results of various
analytic calculations \cite{matveev,go94,gr94,schoell94,sk98},
where a logarithmic divergence for
$F''(n_G$=0.5) was found at $T = 0$. While this divergence cannot be
seen directly in our finite-temperature MC simulations, its precursor
is clearly observed: $F''(n_G$=0.5) increases continuously
as temperature decreases for all $\alpha_t$ values considered here.

\subsection{The charge on the island}

The average number of excess electrons in the island, 
$\langle n \rangle$,  is of
practical interest, since it can be measured directly
by measuring the voltage of the box. 
It follows from the free energy by Eq.\,(\ref{mcharge}).
In Fig.\,7(a) we display $\langle n \rangle$ vs. the gate charge  
$n_G$ at several temperatures for a dimensionless
conductance $\alpha_t = 0.1$.
As expected, at high temperature the average charge follows closely
the line $\langle n \rangle = n_G$, while it decreases
for lower $T$ as Coulomb blockade
effects become more pronounced.
According to the numerical restriction discussed above
for the calculation of the partition function $Z(n_G)$, 
the lowest temperature that could be studied for all values of $n_G$
at $\alpha_t = 0.1$ was $k_B T \sim 0.03 E_C$. 
For these parameters the average charge
$\langle n (n_G)\rangle$ has reached
the zero-temperature limit for $0 < n_G \lesssim 0.35$,
as can be seen from the convergence of the curves in the figure.
 For gate charges $n_G$ closer to 0.5, no saturation
 is found for the temperatures covered,
 and the $T = 0$ value
 will be lower than our finite-temperature results.
 Our conclusions are supported by the
real-time RG calculations by
 K\"onig and Schoeller \cite{sk98}. They find a 
zero-temperature $\langle n (n_G)\rangle$ close to
our low-temperature results up to $n_G \sim 0.35$.
However, consistent with earlier work
\cite{matveev,go94,gr94,schoell94}, the RG calculations predict at $T=0$ a
logarithmic divergence for the slope of $\langle n (n_G)\rangle$ at
$n_G = 0.5$. This is beyond the range covered by the MC data.
We note, however, that the lowest temperature
presented in Fig.\,7(a) is not far from the lowest temperatures
presently attainable in the laboratory, as for typical
values of $E_C/k_B = 1$ K and $T$ = 20 mK, one
has $k_B T = 0.02 E_C$.

\begin{figure}
\centerline{\psfig{figure=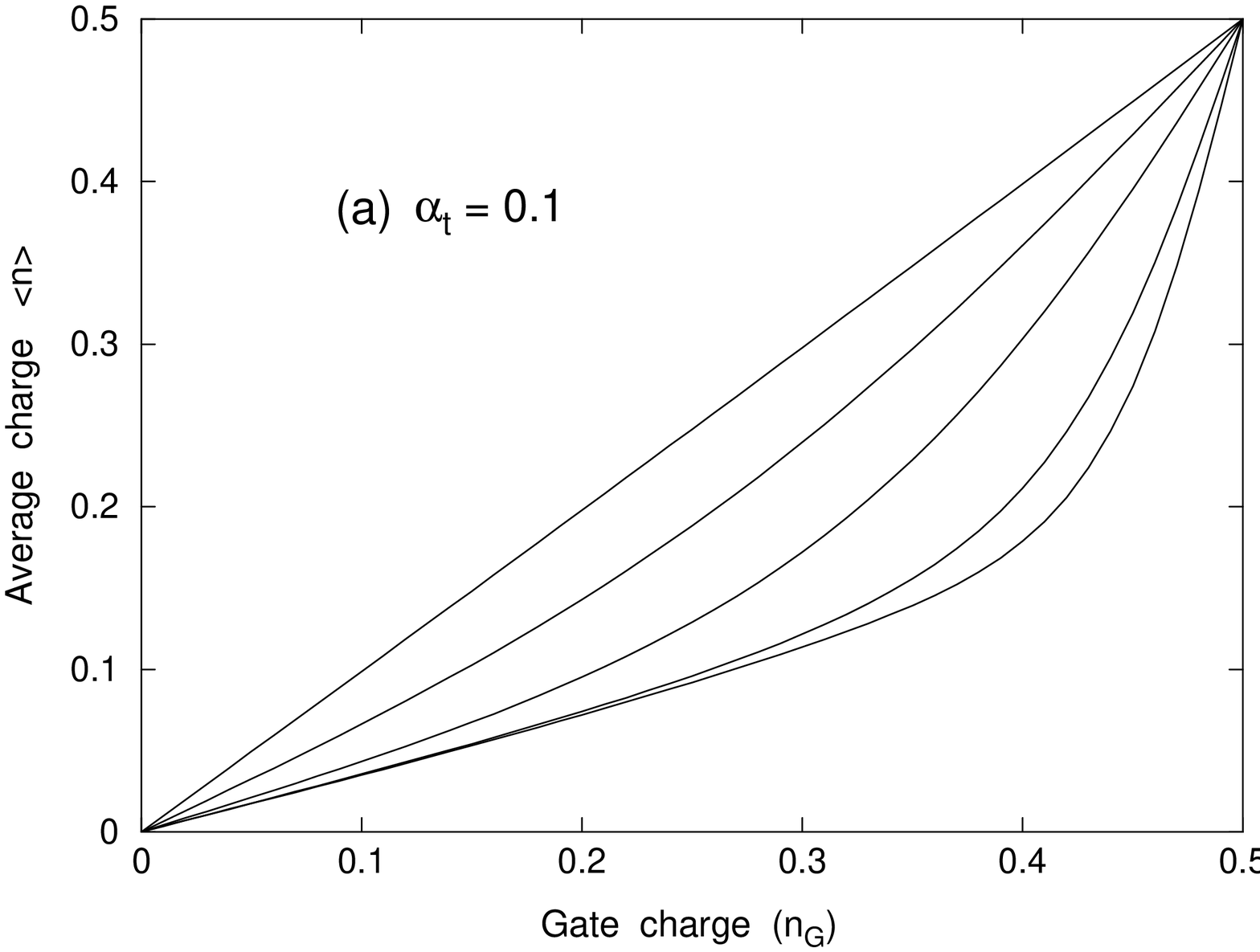,height=7.0cm}}
\vspace*{-1.5cm}
\centerline{\psfig{figure=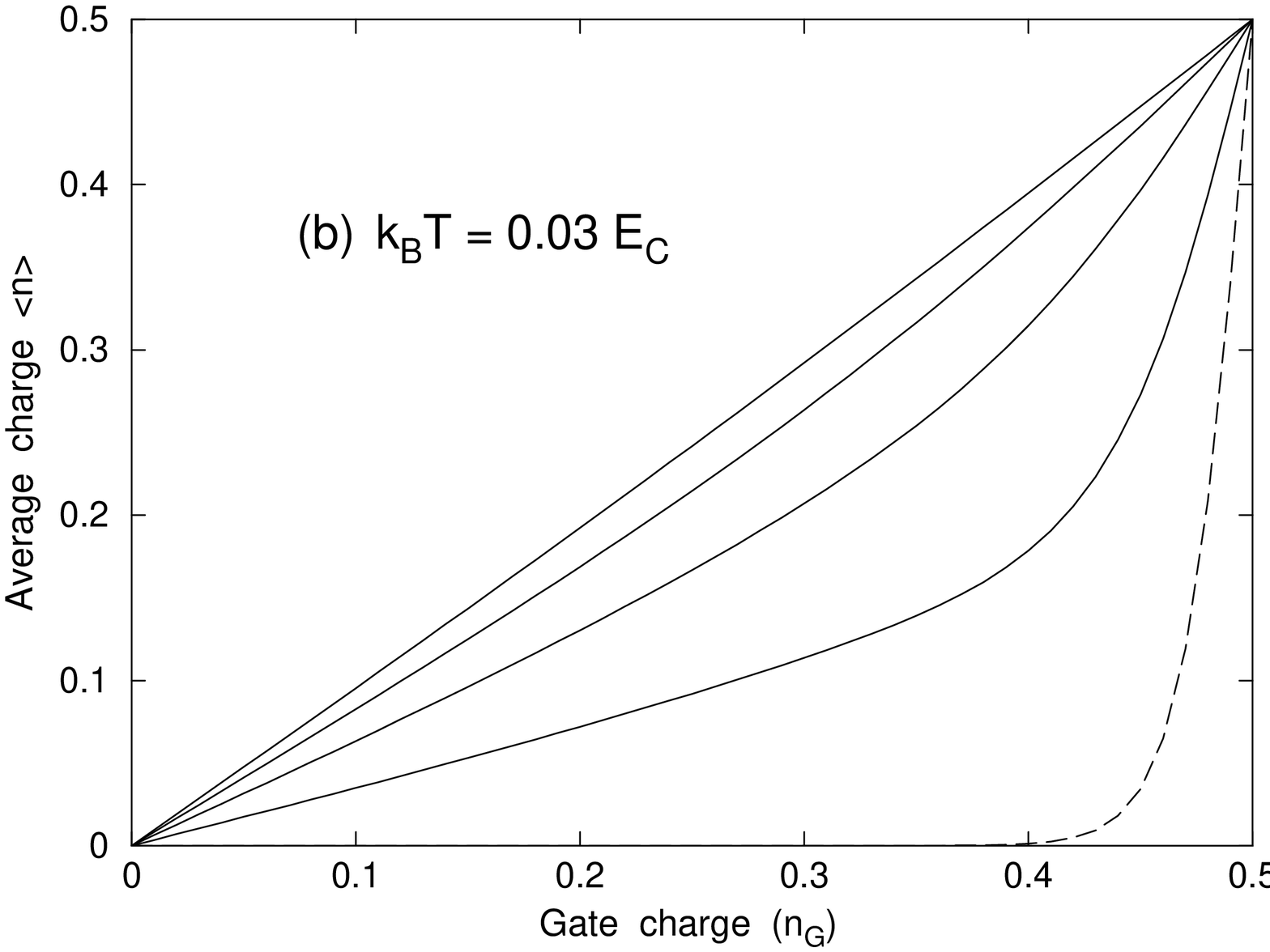,height=7.0cm}}
\vspace*{-1.0cm}
\caption{
Average number of excess electrons, $\langle n \rangle$,
in the single-electron box.
(a) For a dimensionless conductance
$\alpha_t = 0.1$ at various temperatures. From
top to bottom: $k_B T/E_C$ = 0.6, 0.2, 0.1, 0.04, and 0.03.
(b) At  temperature
$T = 0.03 E_C / k_B$
for several values of $\alpha_t$. From top to bottom:
$\alpha_t =$ 0.4, 0.3, 0.2, and 0.1.
For comparison, the average charge for
$\alpha_t = 0$ is also given (dashed line).
}
\end{figure}

\begin{figure}
\centerline{\psfig{figure=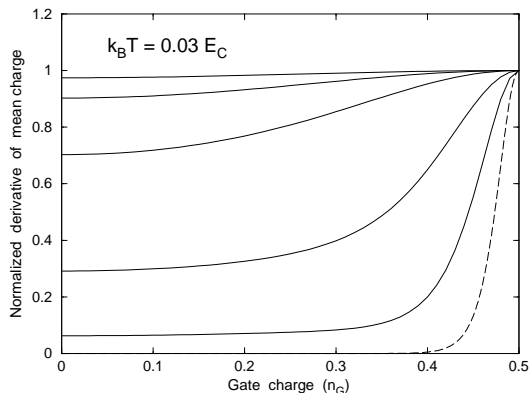,height=7.0cm}}
\vspace*{-1.0cm}
\caption{
Derivative of the mean charge in the island with respect
to the gate voltage,
$\partial \langle n \rangle / \partial n_G$, for
different values of $\alpha_t$ at
$k_B T = 0.03 E_C$.  From top to bottom:
$\alpha_t =$ 0.5, 0.4, 0.3, 0.2,
0.1. The dashed line corresponds to the absence of
electron tunneling ($\alpha_t = 0$).
For each $\alpha_t$, the corresponding curve has been
normalized by the value at $n_G = 0.5$.
}
\end{figure}

In Fig.\,7(b) we present the dependence $\langle n \rangle$ vs.\,$n_G$ for
several values of the tunneling conductance, as obtained
from MC simulations at a temperature $T = 0.03 E_C / k_B$. 
In this figure, the tunneling conductance increases 
from bottom to top, and for the largest $\alpha_t$
shown ($\alpha_t = 0.4$) the dependence is close to a linear one.
For comparison, we give also the average charge for
the case $\alpha_t = 0$ (dashed line).
 These finite-temperature results compare well with
 those found recently from
Monte Carlo simulations in the charge representation 
by G\"oppert {\em et al.}\cite{go98} at a lower
temperature ($\beta E_C = 10^4$).
The main difference between
our finite-temperature results and those reported in ref. \cite{go98}
is found again in the region close to $n_G = 0.5$,
where the above-mentioned logarithmic divergence
for $\partial \langle n \rangle / \partial n_G$ 
is expected to appear.

In order to analyze how long Coulomb blockade 
are observable it is convenient to study the derivative
$\partial \langle n \rangle / \partial n_G$ as a function
of $n_G$. For $\alpha_t \ne 0$ this derivative has its maximum  
at $n_G = 0.5$, and it approaches a plateau around $n_G = 0$, 
the height of which is related to $E_C^*(0)$ by Eq.\,(\ref{ec1}). 
This decrease is a measure for the strength and observability 
of charging effects.
The results of our MC simulations 
are presented in Fig.\,8 for tunneling conductances in the
range $0.1 \le \alpha_t \le 0.5$, at temperature
$k_B T = 0.03 E_C$ ($\alpha_t$ decreases from top to
bottom). For convenience we normalized  
$\partial \langle n \rangle / \partial n_G$ by its value 
at $n_G = 0.5$ for each $\alpha_t$.
We see that even for the highest conductance presented, $\alpha_t = 0.5$,
 charging effects remain observable 
in the range of temperatures comparable to the corresponding 
$E_C^*(0)/k_B$: the decrease of
$\partial \langle n \rangle / \partial n_G$ 
at $n_G = 0$ as compared to the value for $n_G = 0.5$ 
is $\sim 3$\%.
A similar conclusion can be reached for other values of $\alpha_t$, i.e.\ 
charging effects should be observable (at least) at temperatures
of the order of $E_C^*(0)/k_B$  even in the limit of large junction
conductances. These results are compatible with recent
measurements by Chouvaev {\em et al.} \cite{ch98}, where
clear signs of the persistence of charging effects were observed in a sample
with an effective conductance as large as $\alpha_t \approx 0.84$,
even at temperatures larger than the renormalized charging energy.

\section{Discussion}

The detailed MC analysis carried out in the present paper
allows us to  determine the low-temperature values 
of the renormalized charging energy $E_C^*$ for a
single-electron box in the range of weak to strong tunneling
$0 \le \alpha_t \lesssim 0.6$. Our MC data
interpolate well between the perturbative results \cite{gr94} and
those of a nonperturbative instanton analysis \cite{pa91}
in the limits of low and high
$\alpha_t$, respectively. Our data demonstrate that the third order
perturbation theory \cite{go98} yield quantitatively correct values for
$E_C^*$ for $\alpha_t \lesssim 0.25$.
On the other hand, we find quantitative
deviations at higher values of $\alpha_t$.

A similar conclusion has to be drawn concerning the validity
of the RG approach of Ref. \cite{sk98}. This approach is
not equivalent to a direct perturbative expansion \cite{gr94,go98}
in $\alpha_t$ since it allows for a partial 
summation of diagrams in all orders in $\alpha_t$. 
On one hand, for large $\alpha_t$ it works  better than the
perturbative expansion \cite{go98}, capturing the qualitatively correct
trend of the renormalized charging energy $E_C^*$ to decrease with
increasing $\alpha_t$ up to high values. On the other 
hand,  the neglect of higher-order vertex corrections in the RG
may lead to quantitative errors at large  $\alpha_t$.
While the explicit validity range of the RG analysis is difficult to
establish analytically, the corresponding
information can be extracted from a comparison to the MC data. This
comparison 
reveals differences between the RG approach \cite{sk98} and the MC
data for $\alpha_t \gtrsim 0.3$. 

For stronger tunneling we expect a crossover to an exponential 
dependence of $E_C^*$ on $\alpha_t$. This is clearly
shown by our MC data already at $\alpha_t \gtrsim 0.5$, 
where it approaches the result of Ref. \cite{pa91}. 
This crossover is not displayed by the perturbative expansions in powers
of $\alpha_t$, and from a general point of view it appears 
that it cannot be captured within any  finite-order expansion.
For this reason also MC algorithms formulated in the charge
representation may be limited to not too large values of $\alpha_t$.
On the other hand, the real-time RG result \cite{sk98} has been
 fitted to the exponential dependence 
(\ref{EC*}) for $0.5 \lesssim \alpha_t \lesssim 1$, but it requires 
a substantially different pre-exponential function $f(\alpha_t)$
as compared to that found within the instanton approach \cite{pa91}.

Additional support for both the analytical results obtained in Ref. \cite{pa91} 
and our numerical data in the strong tunneling regime comes from an 
independent MC calculation of Ref. \onlinecite{br92}. 
These authors analyzed numerically the correlator
$\langle \cos (\varphi (\tau)-\varphi(0))\rangle$ as a function 
of temperature at different values
of $\alpha_t$, and observed a sharp crossover in the behavior
at a temperature $T^*(\alpha_t)$. Since
this temperature should be close to the renormalized charging
energy $T^*(\alpha_t) \sim E_C^*(0)/k_B$, 
it is interesting to compare both quantities for different values of
$\alpha_t$. This comparison,
carried out in Ref. \onlinecite{za93}, revealed a very
good agreement of the data for $T^*(\alpha_t)$ \cite{br92} with the result
of the instanton calculation (\ref{inst}) (see Fig.\,1 in Ref.
\onlinecite{za93}), which in turn agrees with our MC data in the
strong tunneling regime.

Our numerical data for $E_C^*$ are in agreement with the previous 
MC data of Ref.~\onlinecite{ho97}, extending them to larger $\alpha_t$ and
to substantially lower values of the temperature. 
Although the accuracy of our MC analysis 
is higher as compared to that achieved in Ref. \cite{ho97} (the error 
bars are smaller and the number of points higher), 
it is satisfactory to observe that
our data show the same trend as that found in \cite{ho97}.
Comparing our results to the MC data of Ref. \onlinecite{wa97} we find
good agreement  for low $\alpha_t \lesssim 0.3$, while for larger 
values of $\alpha_t$ their data 
are {\it systematically higher},
clearly showing a different trend with increasing $\alpha_t$ as 
compared to the one indicated by our data. 
There are several reasons why we believe that this discrepancy cannot be 
ascribed to insufficiently low temperatures used
in our simulation:

(i) For all  studied values of $\alpha_t \lesssim 0.6$, the calculated 
$E_C^*(T)$ increases smoothly as the temperature is lowered and 
it clearly saturates at $T \lesssim E_C^*(0)/5$
(see Fig.\,2). The temperature $T=2\times 10^{-3}E_C / k_B$ is 
sufficient to observe saturation for all $\alpha_t$ up to $\sim 0.6$.
Moreover, a systematic difference between our data and those of Ref. 
\cite{wa97} exists already at relatively low tunneling strengths
$0.2 \lesssim \alpha_t \lesssim 0.4$, where the saturation of $E_C^*(T)$
was observed by all groups already at $T \gtrsim 10^{-2}E_C$, i.e.\ well 
above the lowest temperature employed in our simulations.

(ii) The quantity $\tilde E_C(T)$ converges to {\it the same} value 
$E_C^*(0)$, showing saturation at approximately the same 
$T \lesssim E_C^*(0)/5$ as $E_C^*(T)$. This convergence is clearly
seen in Fig.\,3 for $\alpha_t =0.5$. The same behavior of $\tilde E_C(T)$
was observed for all other values of $\alpha_t$. Since the values
$E_C^*(T)$ and $\tilde E_C(T)$ were calculated independently,
the chance of systematic errors is reduced. Furthermore, these
two quantities monotonously converge to the same value 
from below and from above, respectively,  thus 
providing a lower and an upper
bound for $E_C^*(0)$ at each $T$. According to (\ref{ecet2}) these bounds
should merge at sufficiently low temperatures. This
is exactly what our data demonstrate.

(iii) The lowest temperature reached in our MC simulations and in
those of Refs. \cite{wa97,wa96b} is the same. We observe (see e.g. 
Fig.\,3 for $\alpha_t=0.5$) that at higher temperatures 
$k_B T/E_C \sim 0.01$ the data of Wang \cite{wa96b} are still fully
consistent both with \cite{ho97} and with our data. However, the
data point \cite{wa97,wa96b} at a lower $k_B T/E_C \sim 0.002$ is by
more than a factor of 2 higher than all other points. Neither was this
result confirmed by our MC analysis, nor do we see any physical
reason for such a rapid jump of $E_C^*(T)$ at low $T$,
where this quantity should already approach its zero-temperature value.

Very recently one more numerical 
study of the renormalized charging energy in the strong tunneling regime
was performed in Ref. \cite{go98}, using a different MC algorithm. 
The centers of the symbols for the data points \cite{go98}
lie in-between our data and those of Ref. \cite{wa97}. Unfortunately,
the actual error bars are not presented in Ref. \cite{go98} and the 
symbols are so large that they  cover both
our data and those of Ref. \cite{wa97}, thus making a detailed comparison
impossible at this stage.

In summary, the Monte Carlo method employed here has been shown
to be well suited to study charging
effects in the presence of an external voltage at not
too low temperatures. An important advantage of the MC analysis 
is that it covers
both perturbative and nonperturbative regimes and allows us to
describe a crossover between them.
We conclude that the combination of the expansion (\ref{pert}), 
the instanton result (\ref{inst})
and  the numerical data covers all values of $\alpha_t$, thus
providing complete information about the renormalized charging
energy $E_C^*$  at low temperatures. Our results also  resolve
the controversy between the strong tunneling theories \cite{pa91}
and \cite{wa96a} in favor of the former.
The temperature range
for which results for the average charge in the island
$\langle n \rangle$ can be obtained for
$0 \le n_G \le 0.5$, covers most of the
temperature region actually studied in the experiments.  
The Monte Carlo simulations indicate that
charging effects are observable in the single-electron box even for
strong tunneling, at temperatures of the order of
$E_C^*(0)/k_B$.

\acknowledgments
We thank G. Falci, F. Guinea, J. K\"onig and H. Schoeller for 
useful discussions. 
The work was supported by
the Deutsche Forschungsgemeinschaft within SFB 195 and by the INTAS-RFBR
Grant 95-1305. One of us (C.P.H.) acknowledges financial support
from CICYT (Spain) under project PB96-0874.

\end{document}